\documentclass[12pt]{amsart}
\usepackage{mathrsfs}
\usepackage{amssymb} %Extra symbols 
\usepackage{url}
\usepackage{cite}
\usepackage{hyperref}
\usepackage{color}
\definecolor{refcolor}{RGB}{0,0,190}
\hypersetup{
    colorlinks,
    citecolor=refcolor,
    filecolor=refcolor,
    linkcolor=refcolor,
    urlcolor=refcolor
}
\usepackage{array}
\usepackage{booktabs}
\usepackage{graphicx} %Include postscript graphics
\usepackage{amsaddr}

\theoremstyle{definition}

\def\({\left(}
\def\){\right)}

\newcommand{\tn}{\textnormal}
\newcommand{\op}[1]{\tn{#1}}

\newcommand{\R}{\mathbb{R}}

\newcommand{\de}{\tn{d}}

\newcommand{\Diff}{\tn{Diff}}

\newcommand{\vol}{\de_{vol}}

\newcommand{\ds}{\displaystyle}

\newcommand{\ie}{\textit{i.e.} }

\newcommand{\eg}{\textit{e.g.} }
\newcommand{\etc}{\textit{etc.}}

\newcommand{\dsfrac}[2]{\ds{\frac{#1}{#2}}}

\newcommand{\mf}[1]{\mathfrak{#1}}
\newcommand{\mc}[1]{\mathcal{#1}}

\newcommand{\sref}[1]{\S\ref{#1}}

\newcommand{\rank}{\tn{rank }}
\newcommand{\tensors}[3]{\mc T{}^{#1}_{#2}#3}

\newcommand{\idxannih}[2]{#1{}^{#2}{}}
\newcommand{\idxcoannih}[2]{#1{}_{#2}{}}

\newcommand{\annih}[1]{\idxannih{#1}{\bullet}}
\newcommand{\coannih}[1]{\idxcoannih{#1}{\bullet}}

\newcommand{\vectmodule}{\mf X}
\newcommand{\fivect}[1]{\vectmodule(#1)}
\newcommand{\cocontr}{{{}_\bullet}}

\newcommand{\SO}{\op{SO}}
\newcommand{\so}{\mathfrak{so}}

\newcommand{\image}[3]{\begin{figure*}[ht]
\includegraphics[width=#2\textwidth]{#1}
\caption{\small{\label{#1}#3}}\end{figure*}}

\newcommand{\flrw}{Friedmann-Lema\^itre-Robertson-Walker}
\newcommand{\FLRW}{FLRW}
\newcommand{\abs}[1]{|#1|}

\def\hyph{-\penalty0\hskip0pt\relax}
\newcommand{\rstationary}{radical{\hyph}stationary}
\newcommand{\semiriem}{semi{\hyph}Riemannian}

\newcommand{\semireg}{semi{\hyph}regular}

\newcommand{\quasireg}{quasi{\hyph}regular}

\newcommand{\nondeg}{non{\hyph}degenerate}

\newcommand{\nonrenormalizable}{non{\hyph}renormalizable}
\newcommand{\nonrenormalizability}{non{\hyph}renormalizability}
\newcommand{\schw}{Schwarzschild}
\newcommand{\rn}{Reissner-Nordstr\"om}
\newcommand{\kn}{Kerr-Newman}
\newcommand{\hor}{Ho{\v{r}}ava}
\newcommand{\HL}{\hor-Lifschitz}

\makeatletter
\renewcommand\section{\@startsection {section}{1}{\z@}%
                                   {-3.5ex \@plus -1ex \@minus -.2ex}%
                                   {2.3ex \@plus.2ex}%
                                   {\center\normalfont\Large\bfseries}}
\renewcommand\subsection{\@startsection {subsection}{2}{\z@}%
                                   {-3.5ex \@plus -1ex \@minus -.2ex}%
                                   {2.3ex \@plus.2ex}%
                                   {\normalfont\large\bfseries}}
\renewcommand\subsubsection{\@startsection {subsubsection}{3}{\z@}%
                                   {-3.5ex \@plus -1ex \@minus -.2ex}%
                                   {2.3ex \@plus.2ex}%
                                   {\normalfont\bfseries}}
\makeatother

\begin{document} 

\author{Ovidiu Cristinel Stoica}
\address{Department of Theoretical Physics, \\National Institute of Physics and Nuclear Engineering -- Horia Hulubei, Bucharest, Romania.}
\email{cristi.stoica@theory.nipne.ro}

\title[]{Metric dimensional reduction at singularities with implications to Quantum Gravity}

\begin{abstract}
A series of old and recent theoretical observations suggests that the quantization of gravity would be feasible, and some problems of Quantum Field Theory would go away if, somehow, the spacetime would undergo a dimensional reduction at high energy scales. But an identification of the deep mechanism causing this dimensional reduction would still be desirable.
The main contribution of this article is to show that dimensional reduction effects are due to General Relativity at singularities, and do not need to be postulated ad-hoc.

Recent advances in understanding the geometry of singularities do not require modification of General Relativity, being just non-singular extensions of its mathematics to the limit cases. They turn out to work fine for some known types of cosmological singularities (black holes and FLRW Big-Bang), allowing a choice of the fundamental geometric invariants and physical quantities which remain regular. The resulting equations are equivalent to the standard ones outside the singularities.

One consequence of this mathematical approach to the singularities in General Relativity is a special, (geo)metric type of dimensional reduction: at singularities, the metric tensor becomes degenerate in certain spacetime directions, and some properties of the fields become independent of those directions. Effectively, it is like one or more dimensions of spacetime just vanish at singularities. This suggests that it is worth exploring the possibility that the geometry of singularities leads naturally to the spontaneous dimensional reduction needed by Quantum Gravity.

\end{abstract}

%--------------------------------------------------------
% Title and contents

\maketitle

\setcounter{tocdepth}{2}
\tableofcontents

%~~~~~~~~~~~~~~~~~~~~~~~~~~~~~~~~~~~~~~~~~~~~~~~~~~~~~~~~~~~~~~~~~~~~~~~%
\section{Introduction}

{\em Quantum Field Theory} (QFT) and {\em General Relativity} (GR) are the most successful theories in fundamental theoretical physics. Their predictions were confirmed with very high precision, and they seem to offer accurate and complementary descriptions of the physical reality.

Yet, each one of them has some problems, especially when one tries to combine them. GR has the problems of infinities which appear at the {\em singularities}. QFT also has problems with infinities, which appear in the {\em perturbative expansion}, and are usually approached by renormalization techniques. Fortunately, the {\em renormalization group} formalized and solved many of the problems of QFT \cite{SP53RG,GML54RG,BS56RG,BS80intro,Shirkov96,Shirkov99}. The {\em Standard Model} of particle physics is proven to be renormalizable \cite{HV72dimreg,tHooft73dimreg,tHooft98glorious}.

But arguably the greatest difficulties appear when one tries to quantize gravity. General Relativity without matter fields is perturbatively {\nonrenormalizable} at two loops \cite{HV74qg,GS86uvgr}. It requires an infinite number of higher derivative counterterms with their coupling constants. The main reason is the dimension of Newton's constant, which is $[\mathcal G_N]=2-D=-2$ in mass units.

In the quest of understanding the {\em small scale} -- {\em ultraviolet} (UV) limit in QFT, and especially in the approaches to Quantum Gravity (QG), the evidence accumulated so far seems to point in one particular direction. This evidence suggests, or even requires, that there is a {\em dimensional reduction} (sometimes called {\em dimensional flow}) to two dimensions in the UV limit.

In this paper, we review some of the approaches to Quantum Gravity which are based on regularization by dimensional reduction. At this point, it is premature to say which one, if any, is correct, or at least which one is more promising. Our central contribution is to show that some of them are completed, and indirectly endorsed by classical General Relativity, which already is accompanied by dimensional reduction effects, at singularities.

Several distinct approaches aiming to solve the main problems of Quantum Gravity are accompanied by a dimensional reduction. Is the dimensional reduction merely a mark, a side effect appearing in most approaches to Quantum Gravity, or is the main cause of the advances obtained by these approaches? The least we can say is that, if we just introduce in the calculations assumptions of dimensional reduction, the behavior of the perturbative expansions is improved, leading in some cases to renormalizability, or at least to asymptotic safety. Hence, maybe what is important is to introduce the dimensional reduction, and the means to do this may be apparently very different.

This paper will show that several kinds of dimensional reduction which lead to improved behavior of perturbative expansion in other approaches are ensured by the spacetime geometry at singularities in a very concrete way. This builds on our previous work, which showed that the equations at black hole singularities \cite{Sto11e,Sto11f,Sto11g,Sto12e} and Big-Bang singularities \cite{Sto11h,Sto12a,Sto12b,Sto12c} can be expressed, by a sort of change of variables, in terms of finite quantities.

Usually, the apparent incompatibility between QFT and GR which manifests as {\nonrenormalizability} is considered to be the fault of the latter, hence usually the unification proposals start by modifying GR. Various approaches to QG are viewed as a hope which will cure not only the {\nonrenormalizability}, but also the problem of singularities, with the price of giving up one or more fundamental principles of GR.

Here we will take the opposite position: the solution to the problem of singularities comes from GR (by extending the theory at singularities), and it also leads to the desired two-dimensionality in the UV limit, which is needed by Quantum Gravity.

In this paper, we aim to show that the solution to the problem of singularities in GR, developed in \cite{Sto11a,Sto11b,Sto11e,Sto11f,Sto11g,Sto12e,Sto11h,Sto12a,Sto12b,Sto12c} and reviewed briefly in \sref{s_singular_gr}, has implications to Quantum Gravity. Various approaches to QG suggest that if spacetime becomes $2$-dimensional at small scales, the quantization of gravity will become feasible. Some of these hints will be reviewed in Section \sref{s_hints_of_dimensional_reduction}. While dimensional reduction appears to be a desirable ingredient for QG, it would be useful to have an explanation of the reason which led to the dimensional reduction, and a geometric interpretation of its meaning. In Section \sref{s_dimensional_reduction_singularities}, we will explain how benign singularities cause the number of dimensions to be reduced, because of the way the metric becomes degenerate. Then we will try to connect the properties of the dimensional reduction caused by singularities, to those required by some of the approaches to QG.

%~~~~~~~~~~~~~~~~~~~~~~~~~~~~~~~~~~~~~~~~~~~~~~~~~~~~~~~~~~~~~~~~~~~~~~~%
\section{Hints of dimensional reduction coming from Quantum Gravity}
\label{s_hints_of_dimensional_reduction}

The method of regularization through dimensional reduction appeared from the observations that the loop integrations depend on the dimension in a continuous way, so that we can replace the dimension $4$ by $4-\varepsilon$, avoiding the poles, and at the end make $\varepsilon\to 0$ \cite{BG72dimreg,HV72dimreg,tHooft73dimreg}.
The original method of dimensional regularization is rather formal, and apparently without implications to the actual physical dimensions. On the other hand, the fact that Quantum Gravity works fine in two dimensions justifies the consideration of the possibility that at small scales the number of dimensions is indeed reduced. In the following we review some ``signs'' that the spacetime is actually required to become two-dimensional in the small scale limit, while maintaining four dimensions at large scales.

%~~~~~~~~~~~~~~~~~~~~~~~~~~~~~~~~~~~~~~~~~~~~~~~~~~~~~~~~~~~~~~~~~~~~~~~%
%\subsection{Dimensional reduction in Quantum Field Theory}

Suggestions that the two-dimensionality plays an important role appeared in various contexts of QFT. Since the first exactly solvable QFT model was discovered \cite{Thi58}, the two-dimensional QFT proved to lead in a non-perturbative and direct manner to interesting results which can be applied then to make conjectures and find results for four dimensions (see \cite{AAR91} and \cite{frishman2010non} and references therein).

A review of the hints that, in various approaches to QG, a dimensional reduction occurs at small scales, is done by Carlip \cite{Car09SDR,Car10sssst}. Many of these hints involve the {\em spectral dimension}. The spectral dimension was calculated for causal dynamical triangulations (CDT) in \cite{AJL05s}, as evidence showing that the four-dimensional spacetime is recovered at larger scale. This resulted in a trend that various approaches to Quantum Gravity adhered to, consisting in calculating the spectral dimension in the UV limit to see if it is $2$ \cite{LR05fractal,Hor09spectral,Mod2008fractalLQG,Car10sssst}. As it is known (see \eg \cite{SVW2011spectral,SVW2011dispersion,FS2012Axial,Calc2012DiffusionMultiFractional}), while there is a correlation between the spectral dimension and the spacetime dimension, they are not equivalent. While the spectral dimension depends on the spacetime geometry too, it is very different. It represents the effective dimension of the diffusion process, being related to the dispersion relation of the corresponding differential operator. Spectral dimension is a widely used indicator in quantum geometry.

%~~~~~~~~~~~~~~~~~~~~~~~~~~~~~~~~~~~~~~~~~~~~~~~~~~~~~~~~~~~~~~~~~~~~~~~%
%\subsubsection{The renormalization group: asymptotic safety}

General Relativity appears to be perturbatively {\nonrenormalizable}, but the renormalization group analysis may give us useful hints. One possibility is that the infinite number of coupling constants become ``unified'' when approaching the UV fixed point.
In \cite{Wein79AS} S. Weinberg proposed, as solution to the {\nonrenormalizability} of Quantum Gravity, the idea of {\em asymptotic safety}. 
Some evidence accumulated in favor of asymptotic safety \cite{RS01rgqg,Lit04fixed,Nied07asqg,HW05qg,Reu07frg,CAR09qg}, especially near two dimensions \cite{KKN96ren2dim,Lit06fixed}. The spectral dimension near the fixed point appears to be $d_S=2$ \cite{LR05fractal}. In \cite{Nied07asqg} it is shown that the existence of a non-Gaussian fixed point for the dimensionless coupling constant $g_N={\mathcal G}_N\mu^{d-2}$ requires two-dimensionality.

%~~~~~~~~~~~~~~~~~~~~~~~~~~~~~~~~~~~~~~~~~~~~~~~~~~~~~~~~~~~~~~~~~~~~~~~%
%\subsubsection{Causal dynamical triangulations}

In the {\em causal dynamical triangulations} approach \cite{AJL00,AJL04,AJL05r,AJL05s,AJL09aqg}, spacetime is approximated by flat four-simplicial manifolds, similar to quantum Regge calculus \cite{Reg61}. The spacelike edges are taken to be of equal length, and the timelike edges of equal duration. The causality is enforced by requiring a fixed time-slicing at discrete times, and that the timelike edges agree in direction. The path integrals can be calculated non-perturbatively, resulting in four-dimensional spacetimes. The spectral dimension, which is the dimension as seen by a diffusion process, turns out to be four at large distances, but two at short distances \cite{AJL05s}.

%~~~~~~~~~~~~~~~~~~~~~~~~~~~~~~~~~~~~~~~~~~~~~~~~~~~~~~~~~~~~~~~~~~~~~~~%
%\subsubsection{Hints from other approaches}

As pointed out in \cite{Car09SDR,Car10sssst}, there are hints from high temperature string theory \cite{AW88hagedorn} that the thermodynamic behavior becomes two-dimensional at high temperatures. Also, Modesto argued that in Loop Quantum Gravity the effective spectral dimension varies from four at large scales, to two at small scales \cite{Mod2008fractalLQG,Mod2008fractalQST,CM2009fractalSpinFoams}. Other indications of dimensional reduction were reported in \cite{ADFLS10}. Other results concern the spectral dimension in quantum spacetime based on noncommutative geometry \cite{Ben2009fractalQST,MN2010sdQU} and un-gravity \cite{NS2011sdUG}.

We shall see in Section \sref{s_dimensional_reduction_singularities} that the dimensional reduction suggested by some of these approaches is a direct consequence of how metric behaves at singularities.

%~~~~~~~~~~~~~~~~~~~~~~~~~~~~~~~~~~~~~~~~~~~~~~~~~~~~~~~~~~~~~~~~~~~~~~~%
\section{Singularities in General Relativity}
\label{s_singular_gr}

In addition to the problem of quantization, there is another problem which seems to plague General Relativity -- that of singularities. Under general conditions, the evolution equations in GR lead to singularities \cite{Pen65,Haw66i,Haw66ii,Haw67iii,HP70,HE95}. It seems that they are unavoidable. The options seem to be as follows:
\begin{enumerate}
	\item 
give up GR, or at least modify it, and
	\item 
explore the singularities and try to find alternative but equivalent descriptions, which do not have problems with the infinities.
\end{enumerate}

The first approach has been widely explored in the literature.
%begin discrete
For example, one way to avoid the problem of singularities is to consider that spacetime is discrete, as in CDT and LQG. These approaches are very interesting and promising, both for Quantum Gravity and for singularities. Maybe spacetime is indeed discrete, but so far experimental evidence supporting this seems to be absent. Also, on theoretical side, classical General Relativity and Quantum Field Theory, which are the most successful theories, and which are supported by numerous experiments, rely on the hypothesis of a continuous spacetime. Therefore, having a solution to the problems of singularities in classical General Relativity is a good thing, because GR is the most successful theory of gravity. On the other hand, if it is a limit case of a different theory, for instance a discrete one, then our solution to the problem of singularities may indicate the existence of a solution yet to be found in that theory.
%Regarding the problem of singularities, it was believed for long time that it is solved by Ashtekar's ``new variables'', along with solving the problem of quantization. This possibility was suggested by the fact that the variable $\widetilde E^a_i$ defines the metric, which is allowed to be degenerate \cite{ASH87,ASH91,Rom93a}. Unfortunately, even in this case, the connection variable $A_a^i$ may become singular \cf \eg \cite{Yon97}, so this method does not solve the problem of singularities. Then, the LQG approach to singularities became to avoid them.
%end discrete

We focused our research on singularities on the second possibility, which is to find alternative but equivalent formulations of Einstein's equations which do not have problems with the infinities, in classical GR. As we will show in the following, this direction turned out to be very productive, by showing that indeed singularities behave unexpectedly nice in GR. We will review in the following some of these results, because, as the central point of this article is, they may help also taming the infinities occurring when we quantize gravity.

%~~~~~~~~~~~~~~~~~~~~~~~~~~~~~~~~~~~~~~~~~~~~~~~~~~~~~~~~~~~~~~~~~~~~~~~%
\subsection{Singular General Relativity}
\label{ss_singular_gr}

Our initial intention was to construct examples of singularities which can be worked out. We call these {\em benign singularities}. The main property they have is that the metric tensor $g_{ab}$ is smooth, the singular features occurring because $\det g = 0$ at the singular points. This allows the construction of the Christoffel symbols of the first kind
\begin{equation}
	\Gamma_{abc}:=\dsfrac 1 2 \(\partial_a g_{bc} + \partial_b g_{ca} - \partial_c g_{ab}\).
\end{equation}
But $\det g=0$ forbids the construction of the Christoffel symbols of the second kind $\Gamma^c{}_{ab}:=g^{cs}\Gamma_{abs}$, which involves the reciprocal metric $g^{ab}:=(g^{-1})^{ab}$. Consequently, the curvature has to be defined in terms of the Christoffel symbols of the first kind, and not of the second, as it is normal:
\begin{equation}
\label{eq_riemann_curvature_tensor_coord}
	R_{abcd}= \partial_a \Gamma_{bcd} - \partial_b \Gamma_{acd} + \Gamma_{ac\cocontr}\Gamma_{bd\cocontr} - \Gamma_{bc\cocontr}\Gamma_{ad\cocontr}.
\end{equation}

The symbol $\cocontr$ denotes the contraction between covariant indices, and we adopted it because our contraction is more general than the usual one and the index notation would not be appropriate when the metric is degenerate. Normally the contraction between covariant indices also requires the reciprocal metric $g^{ab}$, which becomes singular for $\det g=0$. Luckily, the covariant contraction can be defined in a canonical and invariant way even in the case $\det g=0$, provided that the tensor to be contracted satisfies certain conditions \cite{Sto11a}.

To see how, let us consider for starter two covectors $\omega,\tau\in T^*_pM$, where $M$ is a {\semiriem} manifold and $p\in M$. The contraction of the tensor $\omega\otimes\tau$ is given by $g^{st}\omega_s\tau_t$. This is defined by the metric $g_{ab}$, if $\det g\neq 0$. If $\det g=0$, we can do this only if there are two vectors $u,v\in T_pM$ so that $\omega_a=g_{as}u^s$ and $\tau_a=g_{as}v^s$. In this case, we define the contraction by
\begin{equation}
	\omega_{\cocontr}\tau_{\cocontr} := g_{st}u^s v^t.
\end{equation}
We denote by $\annih{T_p}M$ the subspace of $T^*_pM$ consisting of covectors (or $1$-forms) of the form $\omega_a=(u^\flat)_a:= g_{as}u^s$. We can extend this recipe to more general tensors from $\tensors r s {(T_pM)}$, provided that the components which we contract are from $\annih{T_p}M$. Note that we cannot use it to raise indices, since $u^\flat=(u+w)^\flat$ for any vector $w$ so that $w^\flat=0$. They form the kernel $\ker\flat$, which is zero if and only if $\det g=0$.

From the above considerations we conclude that, to define the curvature as in \eqref{eq_riemann_curvature_tensor_coord}, $\Gamma_{abc}$ has to satisfy, at each point $p\in M$, the condition $\Gamma_{abs}w^s=0$ for any vector $w^s\in T_pM$ which satisfies $g_{st}w^sv^t=0$ for any vector $v\in T_pM$. Metrics satisfying this condition are named {\em \rstationary}, and were studied by Kupeli for the special case when the signature of the metric is constant \cite{Kup87b,Kup96}, and by the author for the general case \cite{Sto11a,Sto11b}.

More details are given, in a manifestly invariant formulation, in \cite{Sto11a}, where we introduced also covariant derivatives for differential forms, and we defined the Riemann curvature tensor \eqref{eq_riemann_curvature_tensor_coord}, for the case when the metric can be degenerate. At such singularities, the Riemann curvature tensor $R_{abcd}$ can be defined in an invariant and canonical way, unlike $R^a{}_{bcd}$. As shown there, a simple condition ensures the smoothness of $R_{abcd}$. Metrics of this type, and the corresponding singularities, are named {\em semi-regular}. From the smoothness of $R_{abcd}$ we obtain, in dimension four, the smoothness of $R_{ab}\det g$ and  $R\det g$, and implicitly of
\begin{equation}
\label{eq_densitized_einstein_tensor}
	G_{ab}\det g:=R_{ab}\det g-\frac 1 2 g_{ab} R\det g.
\end{equation}
This allows us to obtain a smooth densitized version of Einstein's equation
\begin{equation}
\label{eq_densitized_einstein}
	G_{ab}\sqrt{-g}^W + \Lambda g_{ab}\sqrt{-g}^W = \kappa T_{ab}\sqrt{-g}^W,
\end{equation}
where $\kappa:=\dsfrac{8\pi \mc G_N}{c^4}$, and $W= 2$ (in many important cases we can take $W=1$, as we shall see in Section \sref{s_sgr_action}, and as shown in \cite{Sto11h}).
This equation is equivalent to Einstein's equation outside the singularities, but remains smooth at singularities.
%The smoothness condition may be too strong sometimes, but what is important is that the metric is {\rstationary}.

A condition stronger than {\semireg}ity is that of {\em {\quasireg}ity}, which allows a smooth Ricci decomposition of the Riemann tensor
\begin{equation}
\label{eq_ricci_decomposition}
	R_{abcd} = E_{abcd} + S_{abcd} + C_{abcd},
\end{equation}
and leads to another extension of Einstein's equation -- the {\em expanded Einstein equation}, which is tensorial \cite{Sto12a,Sto12b}:
\begin{equation}
\label{eq_einstein_expanded}
	(G\circ g)_{abcd} + \Lambda (g\circ g)_{abcd} = \kappa (T\circ g)_{abcd}
\end{equation}
where the operation
\begin{equation}
	(h\circ k)_{abcd} := h_{ac}k_{bd} - h_{ad}k_{bc} + h_{bd}k_{ac} - h_{bc}k_{ad}
\end{equation}
is the {\em Kulkarni-Nomizu product} of two symmetric bilinear forms $h$ and $k$.

Big-Bang singularities of this type satisfy Penrose's Weyl curvature hypothesis \cite{Pen79,Pen11} automatically \cite{Sto12c}.

A simple but central result in singular {\semiriem} geometry is the generalization, given in \cite{Sto11b}, of {\em warped products} \cite{ONe83}. The warped products allow the construction of a large class of {\semireg} and {\quasireg} singularities. As a corollary, a warped product of (\nondeg) {\semiriem} manifolds is {\semireg}, and in some cases {\quasireg}. The warped product can be used to show that the singularity at the Big-Bang of the {\flrw} model is {\semireg} \cite{Sto11h}, and {\quasireg} \cite{Sto12a}. So it is large enough to include the {\FLRW} singularities, but also the {\schw} singularities \cite{Sto11e}. However, the class of semi-regular singularities is much larger than warped products, as can be seen in \cite{Sto11a,Sto12b,Sto12c}. In addition, more general black hole singularities \cite{Sto11f,Sto11g}, which I do not known yet if they are or not semi-regular, present dimensional reduction effects which will be useful in the following sections.

In \cite{Sto12e} it has been shown how the stationary black hole solutions \cite{Sto11e,Sto11f,Sto11g}, but also solutions which are not eternal and which evaporate, can be foliated in space+time, and are compatible with global hyperbolicity. For the {\rn} solution, foliations were found explicitly \cite{Sto11f}. For the {\kn} solution, it has been shown that the closed timelike curves can be removed by choosing appropriate coordinates \cite{Sto11g}, although this does not mean that singularities eliminate the closed timelike curves in general. 
Since the solutions can be extended analytically beyond singularities, this opens new hopes for the problem of information loss in black holes.

The equations can be extended beyond the singularities in a way which preserves the topology. However, this does not impose any constraints of the topology of the initial spacelike hypersurface. The initial spacelike hypersurface can be any $3$-manifold. This works for warped products such as the {\FLRW} cosmology \cite{Sto11h,Sto12a}, but also for a larger class of solutions, given in \cite{Sto12c}.

%~~~~~~~~~~~~~~~~~~~~~~~~~~~~~~~~~~~~~~~~~~~~~~~~~~~~~~~~~~~~~~~~~~~~~~~%
\subsection{Relation with the first-order formalism}
\label{s_einstein_palatini}

%At a point where the metric is {\nondeg}, there are orthonormal frames, $(e_1,\ldots,e_D)$, so that $g(e_I,e_J)$ is diagonal, and all its diagonal entries are $\pm 1$. Orthonormal frames exist, on a neighborhood, at any point where the metric is {\nondeg}. If the metric is degenerate, there are no orthonormal frames, but there are orthogonal frames, so that $g(e_I,e_J)$ is diagonal. In most cases such frames can be extended locally.

We discuss now the relation with the {\em first order formalism}, also named improperly the {\em Palatini formalism}, although it has been developed by Einstein (see \cite{Ein28,Ein4art,FFR82}). In fact, Palatini's method consists in varying independently the Hilbert-Einstein action with respect to the metric and with respect to the connection. Torsionlessness of the connection is not required from the beginning, but it follows from the variational principle \cite{Pal19}. 
The first order formalism combines Palatini's method with the method of {\em moving frames} developed by Frenet \cite{Fre1852}, Serret \cite{Ser1851}, Darboux \cite{Dar1896}, and Cartan \cite{Car1938}.
If $D=4$, it is sometimes called the {\em tetrad}, or {\em vierbein} method.

Any metric (degenerate or not) on $T_pM$ can be obtained as the pull-back of a {\nondeg} metric $\eta$ on another vector space $\mathbb T_p$, via a linear morphism $e_p:T_pM\to \mathbb T_p$. In most cases, this morphism can be extended to a local bundle morphism $e_U:T_UM\to\mathbb T_U$, where $U\subseteq M$ is an open subset of $M$, and $T_UM$ its tangent bundle. For simplicity, we will consider that $\mathbb T\to M$ is the bundle, and denote the frame bundle morphism by $e:T_UM\to \mathbb T_U$.
In many situations, it is useful to work on the bundle $\mathbb T_U$, and use the bundle morphism $e_U$ to pull back the results on $T_UM$.

If $(x^a)_{a=0}^{D-1}$ is a local coordinate chart on the open set $U\subset M$, and if $(\xi_I)_{I=0}^{D-1}$ is a frame of $\mathbb T\to U$, then 
\begin{equation}
e(\partial_a) = e_a{}^I \xi_I.
\end{equation}
If $X=X^a\partial_a$, we can write
\begin{equation}
e(X)=e(X^a\partial_a)=e_a{}^IX^a\xi_I.
\end{equation}
The metric $g$ on $M$ is defined as the pull-back $g=e^*(\eta)$, that is,
\begin{equation}
g(X,Y)=\eta\(e(X),e(Y)\)
\end{equation}
for any $X,Y\in\fivect M$, and takes the form 
\begin{equation}
\label{eq_metric_frame}
g_{ab}=e_a{}^Ie_b{}^J\eta_{IJ}.
\end{equation}
When $e$ is invertible, if $e^a{}_I:=(e_a{}^I)^{-1}$
\begin{equation}
\label{eq_metric_inverse_frame}
g^{ab}=e^a{}_Ie^b{}_J\eta^{IJ}.
\end{equation}

We consider the frame $(\xi_I)_{I=0}^{D-1}$ to be $\eta-$orthonormal:
\begin{equation}
	\eta_{IJ}=diag(\pm 1,\ldots,\pm 1).
\end{equation}

From now on, we consider that $D=4$, and the metric $\eta$ has signature $(1,3)$.
A connection $\mc D$ on the internal bundle is a {\em Lorentz connection} if it is an $\SO(1,3)$-connection:
\begin{equation}
	\eta(\mc D_Xs,s')+\eta(s,\mc D_Xs')=\eta(s,s')X,
\end{equation}
for any sections $s,s'$ of the internal bundle and vector field $X$ on $M$. In general, a Lorentz connection has the form
\begin{equation}
\label{eq_tetrad_connection}
	\mc D_Xs=\(X(s^I) + A_a{}^I{}_JX^as^J\)\xi_I,
\end{equation}
being therefore defined by the $\so(1,3)$-valued 1-form $A$, named the {\em potential}. Its corresponding {\em curvature} is
\begin{equation}
	F_{ab}{}^{IJ}=\partial_{[a}A_{b]}{}^{IJ} + [A_a,A_b]^{IJ}.
\end{equation}
Because of the skew-symmetry of the elements of the representation of $\so(1,3)$, $A_a{}^{IJ}=-A_a{}^{JI}$ and $F_{ab}{}^{IJ}=-F_{ab}{}^{JI}$. Moreover, $F_{ab}{}^{IJ}$ is skew-symmetric also in the indices $a$ and $b$.

When $e$ is invertible, it allows us to pull-back the Lorentz connection to the tangent bundle, obtaining a connection $\nabla_a$ with Christoffel symbols given by $\Gamma^c_{ab}=e^c{}_Ie_b{}^JA_a{}^I{}_J$, and curvature $R^d{}_{abc}=e^d{}_I e_c{}_J F_{ab}{}^{IJ}$.

If $e$ is not invertible, and the pull-back metric $g$ is degenerate, the curvature $R^d{}_{abc}$ is not defined, because it involves the inverse of $e$, but $R_{abcd}=e_d{}_I e_c{}_J F_{ab}{}^{IJ}$ is well-defined, and coincides with the Riemann curvature defined in equation \eqref{eq_riemann_curvature_tensor_coord}. We will not give the proof for the general case, but since in GR we are concerned only with metrics which are degenerate only on sets which do not contain open subsets, it is easy to see that the identity follows by continuity.

Palatini's action variables are the tetrads $e^a{}_I$ and the Lorentz connections $A_a{}^{IJ}$ on $M$. The metric is viewed as depending on the tetrad field by (\ref{eq_metric_frame}), and consequently the volume form depends on the tetrad as well: $\vol=\vol(e)$.
The {\em Palatini action} in vacuum is 
\begin{equation}
\label{eq_palatini_action_vacuum}
	S_P(A,e):=\frac 1{2\kappa}\int_M e^a{}_I e^b{}_JF_{ab}{}^{IJ}\vol(e).
\end{equation}

By varying this action with respect to $e$ and $A$, one obtains the vacuum Einstein equations. This method can be used as well to obtain the Einstein equations in the presence of matter and a non-zero cosmological constant.

There are several reasons we preferred to develop the formalism described in Section \sref{ss_singular_gr} and \cite{Sto11a,Sto11b,Sto13a,Sto14a}, rather than working only with the first-order formalism.
\begin{itemize}
	\item 
The first-order formalism relies on the existence of the bundle $\mathbb T\to M$. This may be implicit, but the metric $\eta$ is different than $g$, so we have two geometric structures. They are somewhat redundant, since we expect that the geometry of the metric $g$ is obtained from that of $\eta$. 
	\item 
If the metric is degenerate at $p\in M$, $e(T_p)\lneqq \mathbb T_p$, hence $\mathbb T$ is richer than $e(TM)$. This means that there is hidden information in $\mathbb T$, irrelevant to the geometry of $M$.
	\item 
There are more distinct ways in which $\mathbb T\to M$ can be defined globally, to give the same metric $g$.
	\item 
Geometric objects like the connection and curvature on $TM$ appear to be less fundamental, being obtained by pull-back from those on $\mathbb T\to M$. On the other hand, the vanishing of torsion on $\mathbb T\to M$ is either defined via the vanishing of the pull-back torsion on $TM$, or by a variational principle.
	\item 
While the tetrad action variables $e^a{}_I$ can be smooth when the metric is degenerate, the Lorentz connections $A_a{}^{IJ}$ are usually singular. Our approach works in these cases too (see the discussion at the beginning of Section \sref{ss_singular_gr}).
\end{itemize}

The approach developed in \cite{Sto11a} is invariant from geometric viewpoint, does not depend on additional structures other than the metric $g$, has a more direct geometric meaning for the curvature $R_{abcd}$ and covariant derivatives, is more general, and is a direct extension of {\semiriem} geometry, which is more widely used than the first-order formalism in General Relativity. It allowed us to obtain the results reported in \cite{Sto11e,Sto11f,Sto11g,Sto12e,Sto11h,Sto12a,Sto12b,Sto12c}.

Nevertheless, the two approaches complement each other.

%~~~~~~~~~~~~~~~~~~~~~~~~~~~~~~~~~~~~~~~~~~~~~~~~~~~~~~~~~~~~~~~~~~~~~~~%
\subsection{Action principles and singular General Relativity}
\label{s_sgr_action}

Equations \eqref{eq_densitized_einstein} and \eqref{eq_einstein_expanded} are equivalent to Einstein's at the points where the metric is {\nondeg}. This means that, so long as the metric is {\nondeg}, they can be obtained, like Einstein's equations, from the \textit{Hilbert-Einstein action}:
\begin{equation}
\label{eq_hilbert_einstein_action}
	S_{HE-M}(g):=\int_M\Big[\frac 1{2\kappa}(s-2\Lambda)+\mc L_M\Big]\vol,
\end{equation}
whose Lagrangian density is
\begin{equation}
\label{eq_hilbert_einstein_lagrangian}
	\mc L = \dsfrac {1}{2\kappa}(R - 2\Lambda) \vol + \mc L_M \vol,
\end{equation}
where $\mc L_M \vol$ is the Lagrangian density of the matter and $\Lambda$ the cosmological constant.

Equations \eqref{eq_densitized_einstein} and \eqref{eq_einstein_expanded}, like Einstein's, can be derived from the variation of the curvature scalar $R$ and of the determinant of $g$.

The Lagrangian density \eqref{eq_hilbert_einstein_lagrangian} is defined on a larger subset of $M$ than $R_{ab}$ and $R$ (hence than Einstein's tensor). This is because it is possible that $R\sqrt{-g}$ can be smooth even when $R$ is singular, if $\sqrt{-g}$ compensates $R$. In this case, the densitized Einstein tensor \eqref{eq_densitized_einstein_tensor} is non-singular, and it is possible to write a densitized version of Einstein's equation, like \eqref{eq_densitized_einstein}, with $W=1$.

Similarly, in the first-order formalism (see Section \sref{s_einstein_palatini}), if the Riemann curvature tensor of $g$ is the pull-back of the curvature tensor of the connection $A$ from \eqref{eq_tetrad_connection}, then
\begin{equation}
\label{eq_einstein_tensor_density_palatini}
	G^{ab}\sqrt{-g} = \epsilon^{akst}\varepsilon_{IJ}{}^{KL} e^{bI} e^J_k R_{stKL}.
\end{equation}
Equation \eqref{eq_einstein_tensor_density_palatini} allows us to define the Einstein tensor density $G_{ab}\sqrt{-g}$ of weight $W=1$, which is smooth even if the metric $g$ is degenerate, and again the densitized Einstein equation \eqref{eq_densitized_einstein} can be obtained, with weight $W=1$.

This shows that the action principles themselves suggest that Einstein's equation has to admit extensions at singularities.

%~~~~~~~~~~~~~~~~~~~~~~~~~~~~~~~~~~~~~~~~~~~~~~~~~~~~~~~~~~~~~~~~~~~~~~~%
\section{Dimensional reduction at singularities}
\label{s_dimensional_reduction_singularities}

Adopting dimensional reduction only to obtain perturbative quantization of gravity would be an ad-hoc solution. Fortunately, the dimensional reduction emerges from the very structure of singularities.

This suggests the possibility that the solution to the problem of singularities helps understanding the problem of quantization of gravity.

%The standard way to quantize gravity is by perturbative expansion around a background metric (usually taken to be flat):
%\begin{equation}
	%g_{ab} = \eta_{ab} + f_{ab}
%\end{equation}
%where the perturbation is generally considered to be small.

%But if the particles are in fact singularities, then their metric far from being flat. The perturbation expansion cannot be done around a flat metric like $\eta_{ab}$ is normally considered to be, because the metric is in fact degenerate at singularities. It would be like trying to approximate $0$ as $1+\epsilon$.

%~~~~~~~~~~~~~~~~~~~~~~~~~~~~~~~~~~~~~~~~~~~~~~~~~~~~~~~~~~~~~~~~~~~~~~~%
\subsection{The dimension of the metric tensor}
\label{s_dimensional_reduction_singularities_metric}

If the metric tensor $g_{ab}$ is degenerate at a point $p\in M$, then the distance in a part of the directions vanishes. The vanishing directions are given by the isotropic vectors from the vector subspace $\ker(\flat):=T^\perp_p(M)\leq T_pM$, consisting of the tangent vectors which are orthogonal to $T_pM$. These directions can be eliminated if we take the quotient space
\begin{equation}
	\coannih{T_p}M:=\dsfrac{T_pM}{T^\perp_p(M)}
\end{equation}
whose dimension is equal to the rank of the metric at $p$:
\begin{equation}
	\dim{\coannih{T_p}M}=\rank g_p.
\end{equation}
Then,
\begin{equation}
	\annih{T_p}M:=\flat(T_pM)
\end{equation}
has the same dimension as $\coannih{T_p}M$, and in fact they are related by the isomorphism
\begin{equation}
	\flat:\coannih{T_p}M\to\annih{T_p}M
\end{equation}
induced by the morphism
\begin{equation}
	\flat:T_pM\to \annih{T_p}M\leq T^*_pM.
\end{equation}

We see that, because the metric tensor is degenerate, it can be reduced to a metric tensor on the space $\coannih{T_p}M$, and its reciprocal is a metric tensor on $\annih{T_p}M$. The dimension is actually given by the rank of the metric, which can be viewed as the dimension of the metric tensor.
To allow metric contraction, the covariant indices have to ignore the degenerate directions. Also, to admit a well-defined covariant derivative, differential forms have to ignore the degenerate directions. This allows us to define $R_{abcd}$ even though $g^{ab}$ and $\Gamma^a_{bc}$ are singular. Full details about this can be found in \cite{Sto11a}.

Another aspect of dimensional reduction follows from Section \sref{s_einstein_palatini}, where we have seen that {\semireg} geometry can be expressed in terms of the geometry of a connection on a vector bundle $\mathbb T\to M$. When the metric is degenerate, the image of the morphism $e$, from which all information is obtained by pull-back, is of dimension smaller than $D$.

Other aspects of dimensional reduction, more relevant to Quantum Gravity, will be discussed in Section \sref{s_dimensional_reduction_singularities}.

%~~~~~~~~~~~~~~~~~~~~~~~~~~~~~~~~~~~~~~~~~~~~~~~~~~~~~~~~~~~~~~~~~~~~~~~%
\subsection{The measure in the action integral}
\label{s_dimensional_reduction_singularities_measure}

%~~~~~~~~~~~~~~~~~~~~~~~~~~~~~~~~~~~~~~~~~~~~~~~~~~~~~~~~~~~~~~~~~~~~~~~%
\subsubsection{The measure when the metric is degenerate}

When the metric becomes degenerate, its determinant vanishes. Consequently, the volume form
\begin{equation}
\label{eq_volume}
\vol := \sqrt{\abs{\det g}}\de x^0\wedge\ldots\wedge \de x^{D-1} = \sqrt{\abs{\det g}}\de^D x
\end{equation}
tends to $0$ as approaching the degenerate singularities (in non-singular coordinates, in which the metric is smooth).

The action principle is given by
\begin{equation}
\label{eq_action}
S=\int_\mc M\vol(x)\, \mc L
\end{equation}

If the metric is diagonal in the coordinates $(x^\mu)$, then
\begin{equation}
\label{eq_volume_diag}
\vol(x) = \prod_{\mu=0}^{D-1}\sqrt{\abs{g_{\mu\mu}(x)}}\de^D x.
\end{equation}

This shows that, if one allows the metric to become degenerate, one obtains a measure like that studied by G. Calcagni \cite{Calc2010FractalQFT,Calc2010FractalUniverse,calcagni2013multi,calcagni2013quantum}.

%~~~~~~~~~~~~~~~~~~~~~~~~~~~~~~~~~~~~~~~~~~~~~~~~~~~~~~~~~~~~~~~~~~~~~~~%
\subsubsection{Calcagni's fractal universe}
\label{s_dim_red_fractal}

In the {\em fractal universe program} developed by G. Calcagni, the action is taken to be of the form
\begin{equation}
\label{eq_stmes}
S=\int_\mc M\de\varrho(x)\, \mc L(\phi,\partial_\mu\phi),
\end{equation}
with a measure
\begin{equation}
\label{eq_stme}
\de \varrho(x) = v\,\de^D x.
\end{equation}
Initially, Calcagni explored the measure weight of the form
\begin{equation}
\label{eq_stme_sep}
v:=\prod_{\mu=0}^{D-1} f_{(\mu)}(x).
\end{equation}

The original motivation was the hope that this leads to perturbative renormalizability, including for Quantum Gravity \cite{Calc2010FractalQFT,Calc2010FractalUniverse,Calc2011FractalGravity}.
Mathematically, the fractal universe theory relied initially on the Lebesgue-Stieltjes measure (subsequently replaced by fractional measures \cite{Calc2011FractalGeometry}), fractional calculus, and fractional action principles \cite{el2005fractional,el2008fractional,udriste2008euler}.

%~~~~~~~~~~~~~~~~~~~~~~~~~~~~~~~~~~~~~~~~~~~~~~~~~~~~~~~~~~~~~~~~~~~~~~~%
\subsubsection{Singularities and Calcagni's fractal universe}
\label{s_dim_red_singularities_and_fractal}

To obtain the Lebesgue measure from \eqref{eq_stme_sep}, we simply take the following weights:
\begin{equation}
\label{eq_stme_sgr_weights}
f_{(\mu)}(x) = \sqrt{\abs{g_{\mu\mu}(x)}}.
\end{equation}

In terms of $v(x)=\prod_{\mu=0}^{D-1} f_{(\mu)}(x)$, we have
\begin{equation}
\label{eq_stme_v_sgr}
v = \sqrt{\abs{\det g}},
\end{equation}
so that the measure becomes 
\begin{equation}
\label{eq_stme_sgr}
\de \varrho(x) = \sqrt{\abs{\det g}}\de^D x,
\end{equation}
which is just the standard measure from General Relativity, except that in our framework it is allowed to vanish.
This identification makes the results obtained in \cite{Calc2010FractalQFT,Calc2010FractalUniverse} for QFT to be also obtainable as a consequence of the fact that the metric may be degenerate.

The action \eqref{eq_stmes} refers to the Special Relativistic QFT, but the identification \eqref{eq_stme_sgr_weights} we proposed replaces the weights $f_{\mu}$ with the square root of the metric components, which we allowed to tend to zero. This suggests that GR improves QFT at high energies, by reducing the dimension to two. For GR, Calcagni applies the same recipe he used in Special Relativistic QFT: proposes that the weight $v$ multiplies $\sqrt{\abs{\det g}}$ \cite{Calc2010FractalQFT}. This would of course lead to a modified Einstein equation, and modified GR. Our approach  sticks to the standard GR, the only difference being that, in our solution, the measure has this form simply because of the degeneracy of the metric.

Taking all functions $f_{(\mu)}=f$ in \eqref{eq_stme_sep} leads to an isotropic measure, but one can make also anisotropic choices, leading to the breaking of Lorentz invariance. Singularities are also in general anisotropic, because the metric can become degenerate along some of the directions, and remain {\nondeg} along others. This will be discussed in more detail in \sref{s_dimensional_reduction_singularities_lorentz_invariance} and \sref{s_dimensional_reduction_singularities_anisotropy}.

The original argument in \cite{Calc2012FractalGeometryFT} was that this measure leads to perturbative renormalizability, but later, in \cite{calcagni2013quantum}, it was shown that modified measure is not enough to ensure it, and in fact the original power-counting argument from \cite{Calc2012FractalGeometryFT} fails. Yet, in \cite{calcagni2013quantum} is argued that ``[t]he interest in fractional theories is not jeopardized'', and in \cite{calcagni2013multi} that multi-scale theories are still of interest, not only for QG, but also for cosmology (accelerated expansion, cosmological constant, alternative to inflation, early universe, big bounce, \etc).

Maybe the modified measure approach is insufficient to obtain perturbative renormalizability, but the dimensional reduction at singularities is richer and provides some additional features.

%~~~~~~~~~~~~~~~~~~~~~~~~~~~~~~~~~~~~~~~~~~~~~~~~~~~~~~~~~~~~~~~~~~~~~~~%
\subsection{Metric dimension and topological dimension}
\label{s_dimensional_reduction_singularities_metric_vs_topologic}

%~~~~~~~~~~~~~~~~~~~~~~~~~~~~~~~~~~~~~~~~~~~~~~~~~~~~~~~~~~~~~~~~~~~~~~~%
\subsubsection{Metric dimensional reduction and topology}

Let us consider a vector space $V$ of dimension $D$. A symmetric bilinear form $g$ on $V$ defines a scalar product, or a metric. If $g$ is degenerate, $\rank g < D$. The distance vanishes in the directions from $V^\perp$, and the geometric dimension is given by the rank of the metric. If the metric is degenerate, it is smaller than the vector space dimension $D$.

Let us consider now a singular {\semiriem} manifold $(M,g)$. Even if the rank of the metric is at some points lower than $D=\dim M$ (when the metric becomes degenerate), the topological dimension of the manifold remains $D$.

From mathematical viewpoint, in Differential Geometry there are three layers: the {\em topological structure}, the {\em differential structure} and the {\em geometric structure}. The topological structure on the set $M$ is given by an atlas of local charts mapping an open set from $M$ with one from $\R^D$, so that the transition maps are continuous. If the transition maps are differentiable, $M$ becomes a differentiable manifold. If we endow $M$ with a metric tensor, we obtain a geometric structure on $M$. The topological dimension of $M$ is the dimension $D=\dim\R^D$ of the vector space used in the charts of the atlas. The {\em metric dimension}, or the {\em geometric dimension} is given by the rank of the metric, and is allowed to be at most equal to the topological dimension.

The dimensional reduction at singularities is visible also from the degenerate warped products \cite{Sto11b}. The {\em warped product} of two manifolds $(B,g_B)$ and $(F,g_F)$ with {\em warping function} $f$, which is a function on $B$, is defined as the manifold
\begin{equation}
	B\times_f F:=\big(B\times F, \pi^*_B(g_B) + (f\circ \pi_B)\pi^*_F(g_F)\big),
\end{equation}
where $\pi_B:B\times F \to B$ and $\pi_F: B \times F \to F$ are the canonical projections \cite{ONe83}.
We can also write
\begin{equation}
\de s_{B\times F}^2 = \de s_B^2 + f^2\de s_F^2.
\end{equation}
If we allow the warping function $f$ to become $0$, the metric becomes degenerate, but it is {\semireg} \cite{Sto11b}. The multiplication with $f=0$ apparently reduces the manifold $F$ to a point, hence the geometric dimension of $B\times_f F$ is reduced at such points, although the topological dimension remains the same.

In the case when the metric is degenerate of constant signature $(k,l,m)$ and is {\rstationary}, a theorem of Kupeli \cite{Kup87b} shows that the manifold $(M,g)$ is locally isomorphic to a direct product manifold $P\times_0 N$ between a $k$-dimensional manifold $N$ (without a metric, or with metric equal to $0$) and a (\nondeg) {\semiriem} manifold $P$ of signature $(l,m)$. Hence, from the viewpoint of the metric $g$ on $M$, at any point $p\in M$ the $k=D-\rank g$ dimensions associated with the degenerate directions can be ignored locally. We can thus identify the $D$-dimensional manifold $M$ around the point $p$, with the manifold $P$ of dimension equal to $\rank g$. The manifold $(M,g)$ looks locally like a lower-dimensional manifold $(P,h)$. This situation is analogous to that of the gauge degrees of freedom.

If the metric is {\rstationary} and has variable signature, the manifold $(M,g)$ can be identified piecewisely with lower-dimensional manifolds $(P,h)$. The information contained in the metric $g$ of the manifold $M$ can be obtained by pull-back from that of a metric on a manifold $(P,h)$ with smaller topological dimension.

This establishes the connection with the {\em topological dimensional reduction} proposed and studied by D.V. Shirkov and P. Fiziev \cite{shirkov2010coupling,FS2011KG,Fiz2010Riem,FS2012Axial,shirkov2012dreamland}.

%~~~~~~~~~~~~~~~~~~~~~~~~~~~~~~~~~~~~~~~~~~~~~~~~~~~~~~~~~~~~~~~~~~~~~~~%
\subsubsection{Topological dimensional reduction}
\label{s_dim_red_topological}

In \cite{shirkov2010coupling}, D.V. Shirkov studied a $g\varphi^4$ QFT model described by a self-interacting Lagrangian $L=T-V$, where
\begin{equation}
	V(m,g;\varphi) = \dsfrac{m^2}{2}\varphi^2 + \dsfrac{4\pi^{d/2}M^{4+d}}{9}g_d\varphi^4,\,g>0
\end{equation}
To obtain the regularization, D.V. Shirkov worked in a spacetime with variable topology, having a number of dimensions which varies from $D=4$ in the IR limit, to $D=1+d<4$ in the UV limit.
The coupling constant was assumed to run from dimension $D=4$ in the IR regime, to $D=2$ in the UV regime.
Shirkov worked on a toy model obtained by joining two cylinders $S_{R,L}$ and $S_{r,l}$, of radii $R>r$ and lengths $L,l$, with a transition region $S_{coll}$ of varying radius. He obtained a reversal of the running coupling evolution in the two-dimensional region, and the value of the coupling constant gains a finite minimum value $\bar g_2(\infty)<\bar g_2(M_{dr}^2)$, where $M_{dr}$ is the dimensional reduction scale.
The coupling constant decreases in the IR limit as expected, but it has a maximum at $q=M_{dr}$, and in the UV limit decreases again, to the minimum $\bar g_2(\infty)$. 
An interesting possibility envisioned in \cite{shirkov2010coupling} is a novel type of {\em Grand Unified Theory} scenario, where the coupling constants of the Standard Model forces converge without needing an $\op{SU}(5)$ or other leptoquark symmetry.

The idea of topological dimensional reduction is further explored by P. Fiziev and D. V. Shirkov in \cite{FS2011KG,shirkov2012dreamland}, where it is applied to the Klein-Gordon equation, and extended to higher dimensions and multiple variable radii of compactification \cite{Fiz2010Riem}. The next step was to generalize the $(1+2)$-dimensional solution to the non-static case of {\em axial universes}, by allowing the shape function to depend also on time \cite{FS2012Axial}. The solutions to the resulting Einstein equations turned out to be integrable. The obtained dimensional reduction points admit a classification, and there are hopes that they will provide new insights into the nature of the violations of the C, P, and T symmetries \cite{FS2011KG}.

%~~~~~~~~~~~~~~~~~~~~~~~~~~~~~~~~~~~~~~~~~~~~~~~~~~~~~~~~~~~~~~~~~~~~~~~%
\subsubsection{Dimensional reduction: metric or topological?}
\label{s_dim_red_singularities_vs_topological}

It is not clear at this point to what extent our geometric dimensional reduction obtained at singularities is equivalent to the topological dimensional reduction of Shirkov and Fiziev. But what we can say is that in GR there are other fields to be considered in addition to the metric tensor. The information contained in those fields will be, in general, lost by the topological dimensional reduction induced by the metric dimensional reduction.  On the other hand, in order to admit smooth covariant contractions and smooth covariant derivatives (as defined in \cite{Sto11a} for differential forms and tensors with covariant indices), the fields are required to ignore to some extent the degenerate directions. But even under these conditions, they are more general than the fields defined on manifolds of lower topological dimension, and we cannot recover the former from the latter ones by pull-back.

Keeping the points topologically distinct, even when the distance induced by the metric vanishes between them, provides more generality than their identification by topological dimensional reduction.

One important reason to avoid making the topological identification due to the dimensional reduction is that variable topological dimension is not compatible with the foliation of the spacetime in spacelike hypersurfaces. This kind of foliation is important for global hyperbolicity and for avoiding the information loss, as discussed at the end of Section \sref{ss_singular_gr} and in \cite{Sto12e}.

%~~~~~~~~~~~~~~~~~~~~~~~~~~~~~~~~~~~~~~~~~~~~~~~~~~~~~~~~~~~~~~~~~~~~~~~%
\subsection{Metric dimensional reduction and the Weyl tensor}
\label{s_dimensional_reduction_singularities_weyl}

As pointed out in \cite{Car95}, in  dimension lower than four the Weyl curvature vanishes, and the vacuum Einstein equation has only locally flat solutions, or of constant curvature if the cosmological constant is not $0$. This leads to the absence of local degrees of freedom, \ie of gravitational waves for Classical Gravity, and of gravitons for QG. Our approach to the problem of singularities leads unexpectedly to this kind of dimensional reduction and vanishing of the Weyl curvature.

According to a theorem we proved in \cite{Sto12c}, in dimension $4$, the Weyl curvature vanishes at {\quasireg} singularities. The reason is that $\dim{\annih{T_p}M}<D$, because it is equal to the rank of $g$. Recall from \cite{Sto12c} that at {\semireg} (hence at {\quasireg} too) singularities the Riemann tensor lives in $\annih{T_p}M$, hence for $D=4$, it is defined on a space of at most $3$ dimensions, and so is the Weyl tensor $C_{abcd}$. But any tensor having the algebraic properties of the Weyl tensor vanishes in dimension smaller than $4$, hence $C_{abcd}$ vanishes at {\quasireg} singularities. Because at {\quasireg} singularities the Ricci decomposition is smooth, that is, all the terms in \eqref{eq_ricci_decomposition} are smooth, including the Weyl curvature $C_{abcd}$, this means that around the singularity the Weyl tensor remains small. 

Some examples of {\quasireg} singularities, at which therefore the Weyl tensor vanishes, are given in \cite{Sto12a,Sto12b,Sto12c}, and include isotropic singularities, FLRW singularities, and a large class of generalizations of FLRW singularities which are not homogeneous or isotropic, introduced in \cite{Sto12c}. An important example of {\quasireg} singularity is the {\schw} black hole \cite{Sto12b}, which can be used as a classical model for neutral spinless particles. At this time it is not clear whether the singularities of the stationary charged and rotating black holes are {\quasireg}, but they are analytic, the geometric dimensional reduction occurs, and the electromagnetic potential and its field are analytic and finite at $r=0$ \cite{Sto11f,Sto11g}.

The vanishing of the Weyl curvature tensor implies that the local degrees of freedom, \ie the gravitational waves for GR and the gravitons for QG, are absent (see for example \cite{Car95}). As approaching a {\quasireg} singularity, the contribution of the gravitons vanishes. It would be interesting to see how much this helps eliminating the perturbative non-renormalizability of QG.

%~~~~~~~~~~~~~~~~~~~~~~~~~~~~~~~~~~~~~~~~~~~~~~~~~~~~~~~~~~~~~~~~~~~~~~~%
\subsection{Lorentz invariance and metric dimensional reduction}
\label{s_dimensional_reduction_singularities_lorentz_invariance}

When the metric of a spacetime becomes degenerate, it is no longer Lorentzian. The group of transformations which is associated to the metric cannot be a Lorentz group, since the metric is degenerate. The role of the Lorentz group is taken by another group -- the {\em Barbilian group} \cite{Barb39}. But if we factor out the degenerate dimensions, the Barbilian group is reduced to a subgroup, which is a Lorentz group of lower dimension.

The spacetimes with this kind of metric satisfy the Lorentz invariance so long as the metric is {\nondeg}, and when it becomes degenerate, the Lorentz invariance is maintained, but for lower dimensions. One should mention here that this is the best we can do, if we want to include the singularities in the spacetime. Having the full 4-dimensional Lorentz invariance at singularities is not possible, because of the very definition of singularities.

By comparison, other approaches to QG had to give up Lorentz invariance even outside the singularities. It is the case of Loop Quantum Gravity and {\HL} gravity for example, where it is hoped that the four-dimensional Lorentz invariance emerges at large scales.

%~~~~~~~~~~~~~~~~~~~~~~~~~~~~~~~~~~~~~~~~~~~~~~~~~~~~~~~~~~~~~~~~~~~~~~~%
\subsection{Particles lose two dimensions}
\label{s_dimensional_reduction_singularities_charged}

%~~~~~~~~~~~~~~~~~~~~~~~~~~~~~~~~~~~~~~~~~~~~~~~~~~~~~~~~~~~~~~~~~~~~~~~%
\subsubsection{Charged black holes}
\label{s_charged_bh}

There are solutions to Einstein's equation for which the metric has at least one singular component -- let us call such singularities {\em malign singularities}. For example, the stationary black holes apparently are of this type. Luckily, they can be tamed by a coordinate change.

To understand how, let us recall that initially it was considered that the {\schw} solution has a singularity on the event horizon, in addition to that at the ``center'' of the black hole. This opinion lasted until coordinates system in which the metric becomes manifestly finite on the event horizon were proposed \cite{eddington1924comparison,finkelstein1958past}. The metric turned out to be only apparently singular on the event horizon.

Similarly, for the {\schw} \cite{Sto11e}, {\rn} \cite{Sto11f}, and {\kn}  \cite{Sto11g} black holes there are coordinate changes which make the metric analytic at the genuine singularity $r=0$. For example, the {\rn} solution 
\begin{equation}
\label{eq_rn_metric}
\de s^2 = -\left(1-\frac{2m}{r} + \frac{q^2}{r^2}\right)\de t^2 + \left(1-\frac{2m}{r} + \frac{q^2}{r^2}\right)^{-1}\de r^2 + r^2\de\sigma^2,
\end{equation}
characterizing the stationary black holes with electric charge $q$ and mass $m$, the component $g_{tt}\to\infty$ as $r\to 0$. A coordinate change of the form
\begin{equation}
\label{eq_coordinate_ext_ext}
\begin{array}{l}
\left\{
\begin{array}{ll}
t &= \tau\rho^T \\
r &= \rho^S \\
\end{array}
\right.
\end{array}
\end{equation}
transforms the metric to the form
\begin{equation}
\label{eq_rn_ext_ext}
\de s^2 = - \Delta\rho^{2T-2S-2}\left(\rho\de\tau + T\tau\de\rho\right)^2 + \dsfrac{S^2}{\Delta}\rho^{4S-2}\de\rho^2 + \rho^{2S}\de\sigma^2,
\end{equation}
where
\begin{equation}
	\Delta := r^2 - 2m r + q^2 \tn{ (hence $\Delta = \rho^{2S} - 2m \rho^{S} + q^2$)}.
\end{equation}
For $T > S  \geq 1$, the central singularity turns out to be benign \cite{Sto11f}.

These coordinates suggest the possibility that the standard coordinate systems usually employed for the black holes are in fact singular, and the correct coordinates are analytic, like those we have found.

As a bonus, in the new coordinates for the {\rn} and {\kn} metrics, the electromagnetic potential and the electromagnetic fields also become analytic, and they are finite even for $r=0$ \cite{Sto11f,Sto11g}.

The new coordinates allow the spacetime to be foliated. In the {\rn} case, this requires the condition $T\geq 3S$ \cite{Sto11f}. By continuously modifying the parameters characterizing the black holes ($m$, $q$, and $a$), we can construct models in which they appear and disappear by Hawking evaporation. Because of the existence of foliations, we can construct globally hyperbolic spacetimes containing very general singularities \cite{Sto12e}. Thus, singularities do not necessarily destroy information.

%~~~~~~~~~~~~~~~~~~~~~~~~~~~~~~~~~~~~~~~~~~~~~~~~~~~~~~~~~~~~~~~~~~~~~~~%
\subsubsection{Particles as charged black holes}
\label{s_charged_bh_particles}

Although the precise structure of particles is not yet known, classical approximations are known to be useful. From Quantum Mechanics, we know that it is not realistic to consider particles as having a definite trajectory. On the other hand, Feynman's path integral formulation allows the calculation of quantum amplitudes by integrating over all possible classical trajectories. This shows that it is still relevant to understand the properties of particles, from classical viewpoint. But point particles are, from the viewpoint of GR, black holes. This suggests that it is relevant to understand the properties of the singularities occurring in the general relativistic models of particles, even for quantization.

For example, a spinless charged particle can be modeled as a {\rn} black hole.
As we can see from equation \eqref{eq_rn_ext_ext}, in our coordinates, at $\rho=0$ (which is equivalent to $r=0$), the metric loses two dimensions, those corresponding to coordinates $\rho$ and $\tau$. Apparently the metric on the sphere vanishes too, but this only reflects that the warped product describing spherically symmetric solutions to Einstein's equation involves all the concentric spheres, down to $r=0$.

%In addition to dimensional reduction, the light cones degenerate along the time axis \cite{Sto11f}, decoupling the neighboring points and leading to an {\em asymptotic silence} of the type described by Carlip \cite{Car09SDR,Car10sssst}.

Metric dimensional reduction occurs similarly for the {\schw} (describing neutral particles) and the more general {\kn} (describing charged or neutral, spinning or not particles) solutions.

The fact that two dimensions are lost and that the gauge potential and fields remain finite at the singularity are expected to have important impact on field quantization.

One of the major problems of electrodynamics is the fact that the gauge potential and field become infinite as approaching $r=0$. This problem turns out to be removed, by employing our non-singular coordinate system. This sets as one priority in the future developments of our program to see exactly how this affects the perturbative expansions and in the renormalization group analysis.

In the case of neutral particles modeled as {\schw} black holes, the stress-energy tensor is vanishing. For the particles modeled as charged black holes, the only contribution to the stress-energy tensor is given by the electromagnetic field,
\begin{equation}
T_{ab}=F_{a\cocontr}F_{b\cocontr} - \frac 1 4 F_{\cocontr\cocontr'}F_{\cocontr\cocontr'}g_{ab}.
\end{equation}
In the case of {\rn} black holes, 
\begin{equation}
F = q(2T-S)\rho^{T-S-1}\de\tau \wedge\de\rho,
\end{equation}
and it is analytic everywhere, including at the singularity $\rho=0$ \cite{Sto11f}. In the case of {\kn} black holes, the electromagnetic field $F = \de A$ is also smooth, because
\begin{equation}
\label{eq_kn_electromagnetic_potential_smooth}
A = -\dsfrac{q\rho^{S}}{\Sigma(r,\theta)}(\rho^{T}\de\tau + {T}\tau\rho^{{T}-1}\de\rho - a\sin^2\theta\rho^{M}\de\mu),
\end{equation}
where $S\geq 1$, $T \geq S + 1$, and $M\geq S + 1$ \cite{Sto11g}.

Thus, the electromagnetic field and potential are vanishing smoothly at the singularities.

The metric does not determine the topology, hence solutions having non-trivial topologies are possible. For example, building on the idea of Einstein and Rosen to model charged particles as wormholes \cite{ER35}, and on Rainich's results which allow one to obtain, up to a global phase, the electromagnetic field from curvature in an Einstein-Maxwell spacetime \cite{Rai24a,Rai24b,Rai25a,Rai25b}, Misner and Wheeler developed the ``charge without charge'' subprogram of geometrodynamics \cite{MW57,jgf:1962}. In connection with this, in \cite{Sto11f} we have shown how the {\rn} solution can be seen as a charged wormhole whose mouth is degenerate. An open problem regarding modeling the particles as black holes or wormholes is to see what happens with the spin. Including the spin in these models would significantly improve our understanding of the relation between Quantum Mechanics and General Relativity. Proposals of obtaining $\frac{1}{2}$ spin from quantum geometrodynamics are made in \cite{FriedmanAndSorkin1982HalfIntegralSpinFromQuantumGravity,Giulini2009MatterFromSpace}. A model of the electron as a {\kn} black hole is explored in \cite{Burinskii2014GravityShapeSizeElectron} and references therein. A possibility to obtain the rishons \cite{Shupe1979Rishons,Harari1979SchematicModelOfQuarksLeptons,HarariSeinberg1982RishonModel} from topology is presented in \cite{BilsonThompson2005TopologicalPreons,BilsonThompson2007QG-SM}. Also, interesting results in interpreting particles in terms of exotic smooth structures were made in \cite{Asselmeyer2012GeometrizationMatterByExoticSmoothness}.

%~~~~~~~~~~~~~~~~~~~~~~~~~~~~~~~~~~~~~~~~~~~~~~~~~~~~~~~~~~~~~~~~~~~~~~~%
\subsection{Particles and spacetime anisotropy}
\label{s_dimensional_reduction_singularities_anisotropy}

%~~~~~~~~~~~~~~~~~~~~~~~~~~~~~~~~~~~~~~~~~~~~~~~~~~~~~~~~~~~~~~~~~~~~~~~%
\subsubsection{Spacetime anisotropy at singularities}

The metric \eqref{eq_rn_ext_ext} admits a foliation in spacelike hypersurfaces only for $T\geq 3S$ in \eqref{eq_coordinate_ext_ext} \cite{Sto11f}. This condition allows the coordinates to be compatible with the distinction between space and time. But it leads to an anisotropy between space and time, which is manifest when passing to the old {\rn} coordinates $(t,r)$. A rescaling in the coordinates $(\tau,\rho)$ is isotropic, of course in the coordinates 
$(\tau,\rho)$, but the coordinates $(t,r)$ are rescaled anisotropically, due to \eqref{eq_coordinate_ext_ext}.
The diffeomorphism invariance should be considered valid in coordinates $(\tau,\rho)$, but not in the singular coordinates $(t,r)$.

This anisotropic scaling invariance is similar to that which P. {\hor} managed to obtain by modifying the Lagrangian of GR.

%~~~~~~~~~~~~~~~~~~~~~~~~~~~~~~~~~~~~~~~~~~~~~~~~~~~~~~~~~~~~~~~~~~~~~~~%
\subsubsection{Ho{\v{r}}ava-Lifschitz gravity}
\label{s_dim_red_horava}

Inspired by the quantum critical phenomena in condensed matter systems, {\hor}  proposed in 2009 a new model of Quantum Gravity \cite{Hor09qglp}.  
His starting assumption is that the space and time behave differently at scaling -- there is an anisotropic scaling invariance:
\begin{equation}
\begin{array}{l}
\left\{
\begin{array}{ll}
	\mathbf{x}&\mapsto b\mathbf{x}, \\
	t&\mapsto b^z t. \\
\end{array}
\right.
\\
\end{array}
\end{equation}
To describe an UV fixed point, the critical exponent turns out to be $z=D-1=3$, although it is argued that $z=4$ would be even better.
This anisotropy is not required to be a symmetry of the action itself, but of the solutions. The theory describes in the UV limit interacting non-relativistic gravitons, and is power-counting renormalizable in $1+3$ dimensions.
Lorentz invariance is absent in the UV limit, but it is conjectured that it emerges at large distances, where it is hoped that $z\to 1$.
% The resulting field equations are second-order in time, to avoid ghosts. In the same time they are of high order in space, canceling the divergences of the loop integrals. 
The spectral dimension turns out to be two, for high energies, and four for low energies \cite{Hor09spectral}.

The anisotropy breaks the diffeomorphism invariance, and picks out a distinct time direction. This can be expressed in a space$+$time foliation $\mc F$, as in the ADM formalism \cite{ADM62}. The group of diffeomorphisms $\Diff(M)$ reduces to that of diffeomorphisms which preserve the leaves in the foliation.

Some possible inconsistencies, internal and with the observations, in particular concerning the strong coupling and violations of unitarity, are discussed in \cite{CNPS09HL,WSV10HL,Sot11HL,Vis11HL,BPS09HL,KP10HL,HKG10HL,PS10HL,BPS10HL,WW11HL}. Many of these objections arise from the difficulty to prove that GR is recovered in the IR limit.

%~~~~~~~~~~~~~~~~~~~~~~~~~~~~~~~~~~~~~~~~~~~~~~~~~~~~~~~~~~~~~~~~~~~~~~~%
\subsubsection{Singularities and Ho{\v{r}}ava-Lifschitz gravity}
\label{s_dim_red_singularities_vs_horava}

While the anisotropy $T\geq 3S$ we obtained in \cite{Sto11f} resembles that from {\HL} gravity, there are important differences, because ours follows from standard GR considerations, while that of {\hor} from modifications of the Einstein-Hilbert Lagrangian (and implicitly of Einstein's equation).

In our proposal, the metric is still the fundamental field. The very degeneracy of the metric imposes conditions on the foliation at the singularity. There is no need for other structure to define the foliation, as it is in {\HL} gravity. In our approach there is no need to impose in the IR limit the recovery of standard GR, since we start from standard GR outside the singularities, and extend it in the singularities.

%~~~~~~~~~~~~~~~~~~~~~~~~~~~~~~~~~~~~~~~~~~~~~~~~~~~~~~~~~~~~~~~~~~~~~~~%
\subsection{Does dimension vary with scale?}
\label{s_dimensional_reduction_singularities_scale}

So far we provided arguments that the singularities characterized by the degeneracy of the metric explain the geometric dimensional reduction. This dimensional reduction has many common features with the dimensional reduction expected in other proposals in the literature. In addition, it is not invented with the problem of quantization in mind, but it is a consequence of our approach to the singularity problem, which in turn fits naturally in classical GR.

Yet, the hardest part remains to be done. The geometric dimensional reduction we propose becomes manifest as the distance to a singularity becomes smaller. But the dimensional reduction needed in QFT and QG seems to have nothing to do with the distance to a singularity. It is required just to depend on the scale.

The precise ways in which the geometric dimensional reduction we proposed impacts QFT and GR need to be analyzed in more depth.

It is true that at the points where the metric is regular, usual methods show that gravity is perturbatively non-renormalizable, and QFT has the same problems and properties. On the other hand, the usual perturbative expansion considers that the particles only perturb the metric around a flat metric. But if particles are, when viewed as point particles, singularities, then we should take this into account in the path integrals. Virtual particles, hence virtual singularities, are present at any point of spacetime. Thus, even though classical particles may be singularities, the path integral may be without singularities, and yet, the dimensional reduction contributes to the integral.

The higher order Feynman diagrams involve a larger number of particles. This means that, the higher the order of the Feynman diagram, the larger the number of particles in the same region of space, which we will consider to be benign singularities. Because of this, the metric will have, in average, smaller determinant, and smaller Weyl curvature $C_{abcd}$. Recall that for benign singularities both the determinant of the metric and the Weyl curvature tend to $0$ as the distance to the singularity decreases, and having a higher number of singularities in the same region reduces, in average, these quantities (fig. \ref{dim-red-feynman}). Thus, in the high energy limit, the dimensional reduction will become more and more present in the integrals.

\image{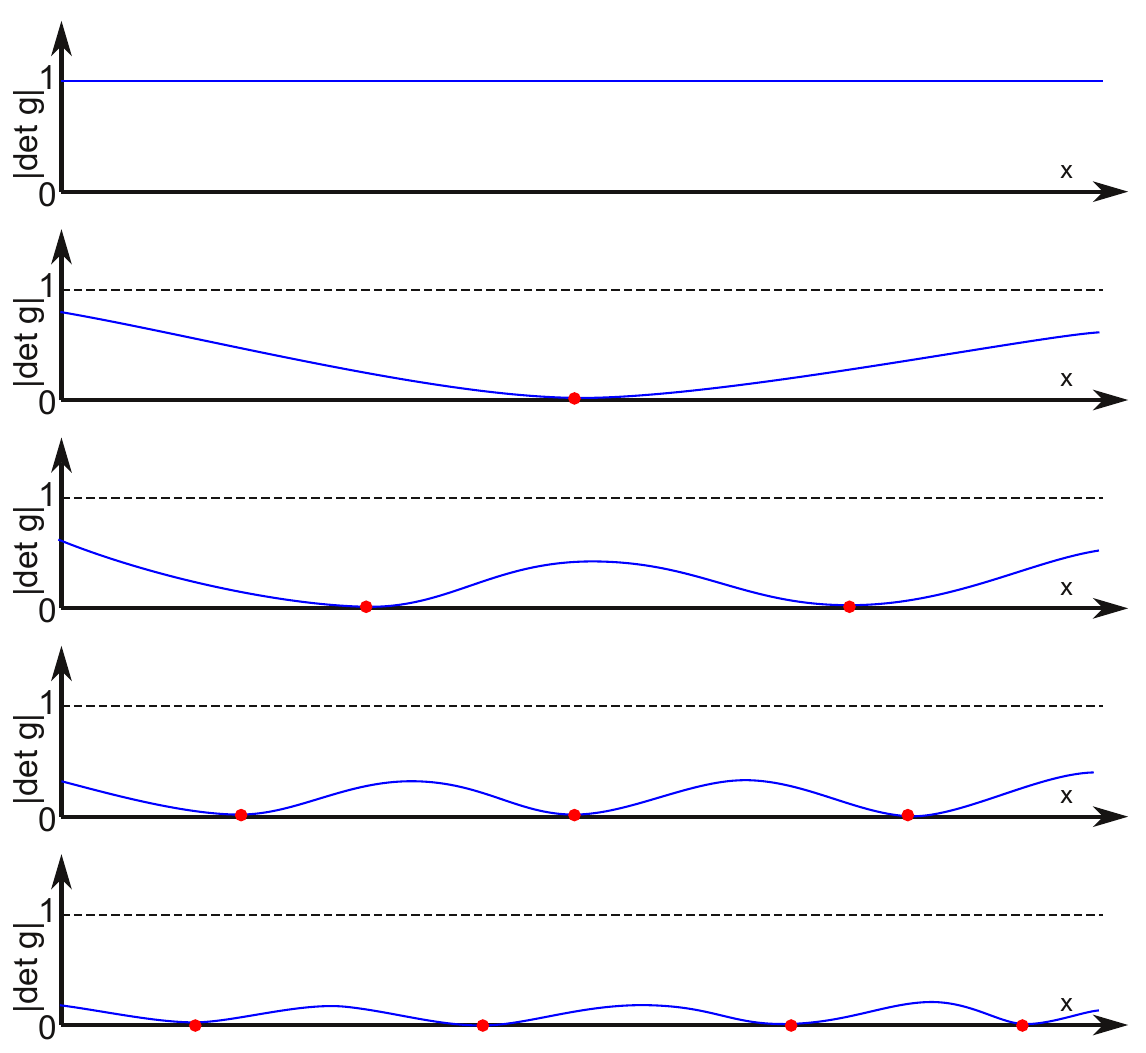}{0.7}{Schematic picture illustrating how we expect that the metric's average determinant decreases as the number of singularities (\ie particles) in the region increases. The red dots represent the singularities, and the blue line represents $|\det g|$.}

In conclusion, \textit{the main conjecture} is that, although the dimensional reduction happens at singularities which represent the particles, when many particles are present, the measure dimensionality and the Weyl tensor are reduced in average, and vanish asymptotically. We should consider the possibility that this acts like a regulator. Let us name this \textit{the hypothesis of average dimensional reduction}.

%~~~~~~~~~~~~~~~~~~~~~~~~~~~~~~~~~~~~~~~~~~~~~~~~~~~~~~~~~~~~~~~~~~~~~~~%
\section{Conclusions}
\label{s_conclusions}

We reviewed some of the hints indicating that if a dimensional reduction would take place at small scales, then some major problems concerning the quantization of gravity, but also of other fields, would go away. Some hints refer to the dimension involved in calculations, others to the geometric and topological dimensions, and others to the spectral dimension.

We advocated here the position that the approach to singularities introduced and developed in \cite{Sto11a,Sto11b,Sto11e,Sto11f,Sto11g,Sto12e,Sto11h,Sto12a,Sto12b,Sto12c}, leading to a (geo)metric dimensional reduction, also opens new perspectives on the quantization of gravity by perturbative methods. This position is supported by the strong connections between the metric dimensional reduction and the other kinds of dimensional reduction, reviewed in this paper. 

This is just a small step; many questions remain open, and much work remains to be done.

%--------------------------------------------------------
\section*{Acknowledgements}
The author thanks P. Fiziev and D. V. Shirkov for helpful discussions and advice received during his stay at the Bogoliubov Laboratory of Theoretical Physics, JINR, Dubna. The author thanks an anonymous reviewer, whose suggestions significantly improved the quality and completeness of this article.

%\bibliographystyle{plain}%{unsrt}%{amsalpha}%{amsplain}
%\bibliography{../bib/sing-gr_bib}

\end{document}